\documentclass[twocolumn, a4paper, 10pt, showpacs, 
reprint,prb, superscriptaddress, nobibnotes, amsmath, 
amssymb, amsfonts
]{revtex4-2}

\usepackage{amsmath}
\usepackage{graphicx}
\usepackage{dcolumn}
\usepackage{bm}
\usepackage{hyperref}
\usepackage{xcolor}
\usepackage{makecell}

\begin{document}
\title{Tuning flat bands by interlayer interaction, spin-orbital coupling, and external fields in twisted homotrilayer MoS$_2$}
\author{Yonggang Li}
\affiliation{Key Laboratory of Artificial Micro- and Nano-structures of Ministry of Education and School of Physics and Technology, Wuhan University, Wuhan 430072, China}
\author{Zhen Zhan}
\email{Corresponding author: zhenzhanh@gmail.com}
\affiliation{Imdea Nanoscience, C/ Faraday 9, 28015 Madrid, Spain}
\affiliation{Key Laboratory of Artificial Micro- and Nano-structures of Ministry of Education and School of Physics and Technology, Wuhan University, Wuhan 430072, China}
\author{Shengjun Yuan}
\email{Corresponding author: s.yuan@whu.edu.cn}
\affiliation{Key Laboratory of Artificial Micro- and Nano-structures of Ministry of Education and School of Physics and Technology, Wuhan University, Wuhan 430072, China}
\affiliation{Wuhan Institute of Quantum Technology, Wuhan 430206, China}
\date{\today}
	
\begin{abstract}		
Ultraflat bands have already been detected in twisted bilayer graphene and twisted bilayer transition metal dichalcogenides, which provide a platform to investigate strong correlations. In this paper, the electronic properties of twisted trilayer molybdenum disulfide (TTM) are investigated via an accurate tight-binding Hamiltonian. We find that the highest valence bands are derived from the $\Gamma$-point of the constituent monolayer, and they exhibit a graphene-like dispersion or become isolated flat bands that are dependent on the starting stacking arrangements. The lattice relaxation, local deformation, and external fields can significantly tune the electronic structures of TTM.  After introducing the spin-orbital coupling effect, we find a spin-valley-layer locking effect at the minimum of the conduction band at the $K-$ and $K^\prime$-point of the Brillouin zone, which may provide a platform to study optical properties and magnetoelectric effects. 
\end{abstract}
	
\maketitle

\section{Introduction} 
Since the discovery of single-layer graphene, research on two-dimensional (2D) materials has drawn significant attention in the scientific community~\cite{novoselov2004electric}. Stacking 2D materials with the rotation angle or lattice mismatch between layers, moir{\'e} superlattices with periodicity that ranges from nanometers to micrometers can be formed \cite{bistritzer2011moire,geim2013van,foo2023extended}. Rotation angle and lattice mismatch , as degrees of freedom, could tune the electronic structures of the moir{\'e} systems. In twisted bilayer graphene (TBG), when the rotation angle approaches 1.05$^\circ$, the so-called magic angle, two van Hove singularities (VHSs) in the density of states (DOS) merge in the charge-neutrality point, resulting in a sharp peak associated with flat bands \cite{kuang2021collective, bistritzer2011moire}. In such a flat band system,  exotic phenomena, including unconventional superconductivity \cite{sharma2022superconductivity, cao2018unconventional}, strong correlations \cite{cao2018correlated}, the quantum anomalous Hall effect \cite{serlin2020intrinsic}, ferromagnetism \cite{sharpe2019emergent}, and electronic collective excitations \cite{kuang2021collective, hesp2021observation}, have been observed, which are not observed in the parent material.
	
Flat bands are also detected in many other moir{\'e} systems, for example, and twisted hexagonal boron nitride \cite{walet2021flat}, twisted bilayer transition metal dichalcogenides (tb-TMDs) \cite{naik2018ultraflatbands,zhang2020flat}. In fact, both the structural and electronic structures of tb-TMDs are remarkably different from that of TBG: Firstly, the appearance of flat bands is different. Flat bands are observed in tb-TMDs with twist angles below 7$^\circ$ \cite{wu2018hubbard,wu2019topological,naik2020origin,venkateswarlu2020electronic, jin2019observation, regan2020mott, tang2020simulation, wang2020correlated, zhang2020flat}, but only special angles in TBG \cite{bistritzer2011moire}. Secondly, 
the localization of the flat band states in real space is different. In TBG, the flat band states at the $K-$point of the Brillouin zone (BZ) are concentrated in the AA region \cite{guinea2018electrostatic}. The tb-TMDs have two distinct configurations, and the localization of the flat band states at the $\Gamma$-point of the BZ is different \cite{naik2018ultraflatbands,vitale2021flat}. Thirdly, the lattice relaxation changes differently the distribution of stackings and interlayer spacings in the two different configurations. Finally,  tb-TMDs are a new model system to explore quantum phenomena, for example, moir{\'e} excitons \cite{andersen2021excitons,tran2020moire, tran2019evidence, sun2022enhanced, seyler2019signatures, shimazaki2020strongly, zhang2023every}, Wigner crystal \cite{regan2020mott}, pair density waves \cite{zhang2020moire}, and plasmons \cite{kuang2022flat}.

Multilayer graphene moir{\'e} systems also possess flat bands, for example, twisted trilayer systems and even twisted multilayer superlattices \cite{cea2019twists}. Different from the twisted bilayer, twisted multilayer systems are more flexible and controllable, which can be easily tuned by twist angles, starting stacking arrangements, and external fields \cite{wu2021lattice}. Furthermore, the moir{\'e} of moir{\'e} twisted trilayer graphene (TTG) has an extended magic phase \cite{foo2023extended,popov2023magic}, and electric field-tunable superconductivity was observed in mirror-symmetry TTG \cite{hao2021electric,cao2021pauli}. Recently, several groups have started to explore the electronic structures of twisted multilayer TMDs. For example, twisted trilayer $\rm WSe_2$ with two different twist angles is successfully fabricated in experiment, which reveals multiple moir{\'e} exciton splitting peaks due to the presence of deeper moir{\'e} potential \cite{zheng2023exploring}. The ABBA-stacked twisted double bilayer WSe$_2$ could be served as a realistic and tunable platform to simulate a $\Gamma$-valley honeycomb lattice with both sublattice and SU(2) spin rotation symmetries \cite{pan2023realizing}. The ABAB-stacked double bilayer WSe$_2$ as a platform to study electronic correlations within the $\Gamma$-valley moir\'e bands, could have control over the spin and valley character of the correlated ground and excited states via the electromagnetic fields \cite{foutty2023tunable}. However, detailed knowledge of the electronic structures of twisted multilayer transition metal dichalcogenides is still missing. 

In this paper, we use a tight-binding (TB) model to study the electronic properties of twisted trilayer molybdenum disulfide (TTM) with different starting stacking arrangements. 
Similar to TTG, the electronic structures of TTM depends strongly on the starting stacking arrangements. By modulating the arrangement, isolated flat bands could be achieved in the valence-band edge.  
Previous investigations show that lattice relaxation has a significant impact on the electronic properties of TBG and tb-TMDs \cite{zhang2021electronic, kuang2022flat, zhan2020tunability}. We find that the lattice relaxation has a distinct influence on the structural and electronic properties of TTM with different configurations due to the interplay between different bilayer moir\'e patterns. Furthermore, these configurations respond differently to the external electric and magnetic fields. By manipulating these degree of freedoms, for instance, the arrangements, strains, spin-orbital coupling (SOC), the electric and magnetic fields, we could tune the flat bands in TTM, creating a promising platform for exploring the strongly correlated states. Different from the tb-TMDs, due to the spin-valley-layer locking in the TTM, the spins could be manipulated by the electric field control of the layer polarization and magnetic field control of the valley polarization, which may lead to new optical and magnetoelectric effects. 

The paper is organized as follows: In Sec. \ref{methods}, we introduce the TTM structure and the numerical methods. In Sec. \ref{sec3}, we investigate the flat bands in TTM with different starting stacking arrangements, and we tune the flat bands with the interlayer interaction, spin-orbital coupling, and external electric fields. Finally, we give a summary in Sec. \ref{conclusion}.

\section{Methods} 
\label{methods}
In TBG, the AA and AB starting stacking arrangements give exactly the same band structures, whereas the electronic structures in twisted trilayer graphene are strongly dependent on the starting stacking arrangements \cite{wu2021lattice,li2019electronic}. In the bilayer TMDs, unlike graphene, the inversion symmetry is present in AB (2H) stacking but broken in AA (3R) stacking. The AB twisted TMDs have AB regions that form a honeycomb network, and the AA twisted one has the AA regions that form a triangular lattice. Consequently, these two distinct moir{\'e} systems show different electronic properties \cite{naik2018ultraflatbands,zhan2020tunability,zhang2020flat}. That is, the bandwidth and localization of the flat bands are significantly different in these two distinct tb-TMDs. The starting stacking arrangements may also play a significant role in the twisted trilayer MoS$_2$. Moreover, in the TTM that is composed of two bilayer moir\'e patterns, there is an interplay between these two patterns in both structural and electronic structures, which may provide new effects in the TTM.
\subsection{The atomic structures}
\begin{figure}[t!]
    \centering
    \includegraphics[width=0.5\textwidth]{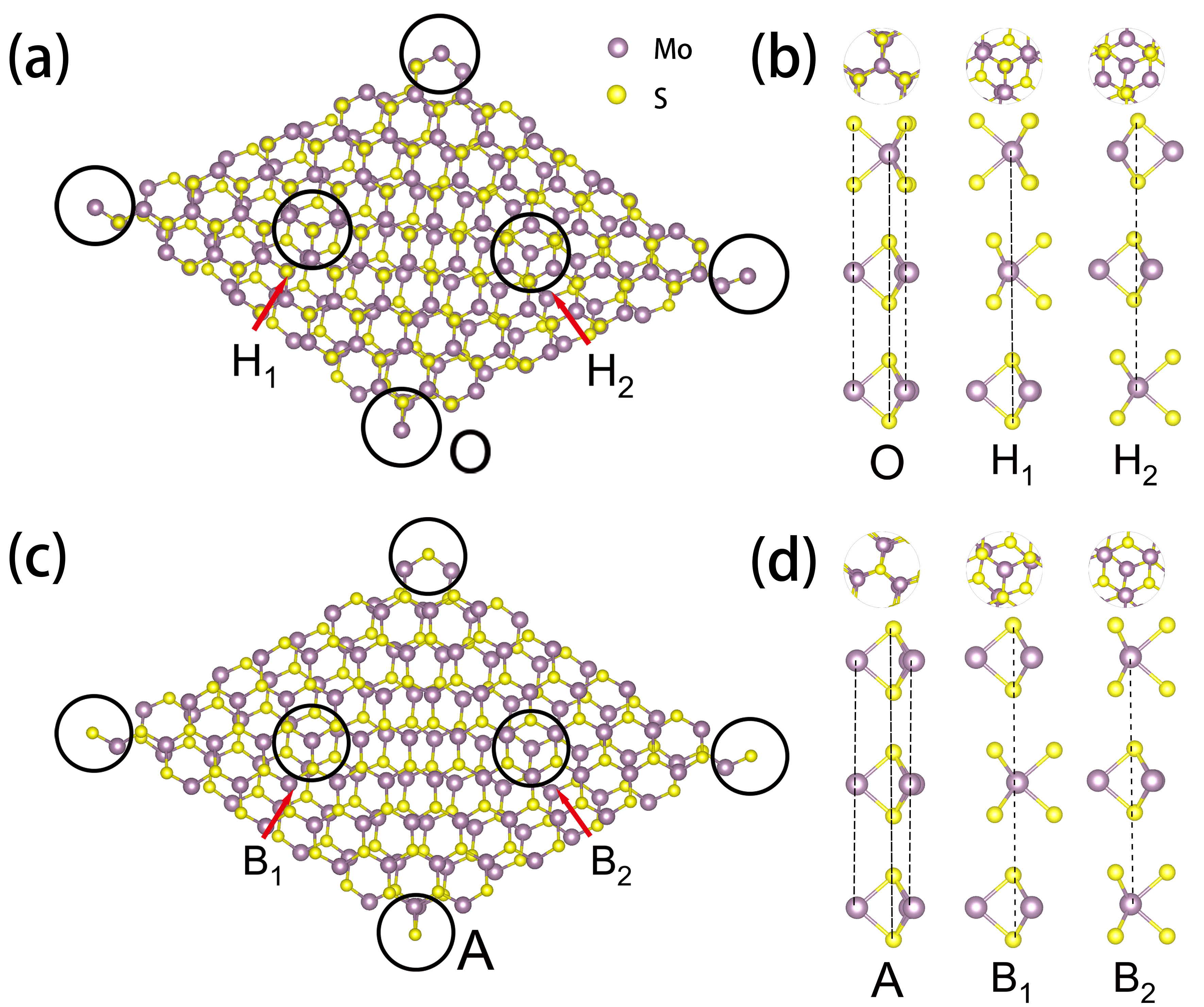}
    \caption{The atomic structure of twisted trilayer MoS$_2$. Parts (a) and (c) are the top views of $\rm A\Tilde{A}B$-7.34$^\circ$ and $\rm A\Tilde{A}A$-7.34$^\circ$. The high-symmetry stackings are highlighted by circles. (b) The stackings at O, H$_1$, and H$_2$ are AAB, S-Mo-Mo, and Mo-S-S, respectively (labeled from bottom to top). (d) The stackings at A, B$_1$, and B$_2$ are AAA, S-Mo-S, and Mo-S-Mo, respectively.}
    \label{fig:structure}
\end{figure}

 In this paper, we focus on two different starting stacking arrangements, AAB and AAA. Based on these two trilayer stackings, we rotate the middle layer to construct a commensurate TTM. Note that only one moir{\'e} pattern is formed in the atomic structure here. That is, the two bilayer patterns have the same moir\'e periodicities, which is different from the moir{\'e}-of-moir{\'e} structures \cite{foo2023extended}. Therefore, similar to the bilayer case, the basis vectors in the TTM can be expressed as
\begin{align}
\vec{t}_1 &= -m\vec{a}_1+(m+n)\vec{a}_2, \nonumber \\
\vec{t}_2 &= (m+n)\vec{a}_1-n\vec{a}_2,
\label{equation:vectors_sc}
\end{align}
where $\vec{a}_1=(a/2,\sqrt{3}a/2)$, $\vec{a}_2=(-a/2,\sqrt{3}a/2)$ are the vectors of monolayer $\rm MoS_2$, with $a = 0.316$ nm being the lattice constant, and $m$ and $n$ are positive prime integers with $n-m=1$, which means that the moir{\'e} supercell contains only one moir{\'e} pattern. Another two parameters to determine the structure of $\rm MoS_2$ are the vertical distance  between sulfur atoms within a monolayer, $\rm d_{S-S} = 0.317$ nm and the vertical distance between the layers, $c = 0.6145$ nm. The twist angle is $ \cos\theta = \frac{m^2+4mn+n^2}{2(m^2+mn+n^2)}$, and the number of atoms in each moir{\'e} supercell is $N=9(m^2+mn+n^2)$. In our calculation, the integer pair $(m ,n)=(10,11)$ gives a twist angle $\theta=3.15^{\circ}$, and one unit cell contains 993 $\rm Mo$ atoms and 1986 $\rm S$ atoms. We use the notations $\rm A\Tilde{A}B$-$3.15^\circ$ and $\rm A\Tilde{A}A$-$3.15^\circ$ to denote these two configurations. The number 3.15 represents the twist angle, and the twisted middle layer is marked by a tilde above the letter. Figure \ref{fig:structure} shows the atomic structure of $\rm A\Tilde{A}B$ and $\rm A\Tilde{A}A$, in which the atoms inside the parallelogram form a unit cell. There are three different types of high-symmetry points in each unit cell, that is, the O, H$_1$, and H$_2$ high-symmetry stackings in $\rm A\Tilde{A}B$, shown in Fig. \ref{fig:structure}(b) and the A, B$_1$, and B$_2$ high-symmetry stackings in $\rm A\Tilde{A}A$, shown in Fig. \ref{fig:structure}(d). 

	
\subsection{Tight-binding model}
We employ a TB model to calculate the electronic properties of TTM~\cite{zhan2020tunability,fang2015abinitio}. The Hamiltonian of the TB model can be expressed as
\begin{equation}
\hat{H} = \hat{H}^{mono}_1 + \hat{H}^{mono}_2 + \hat{H}^{mono}_3 + \hat{H}^{int}_{12}  + \hat{H}^{int}_{23},
\label{equation:hamiltonian}
\end{equation}
where $\hat{H}^{mono}_{1,2,3}$ are the Hamiltonian of monolayer $\rm MoS_2$, and $\hat{H}^{int}_{12,23}$ are the interlayer interactions of bottom-middle layers and middle-top layers, respectively. The primitive cell of monolayer $\rm MoS_2$ consists of one Mo atom and two S atoms, and the relevant atomic orbital basis includes five $d$ orbitals of each Mo atom and three $p$ orbitals of each S atom. In the monolayer Hamiltonian, we only consider the interaction terms between orbitals of the same type at first-neighbor positions, and terms between orbitals of different types at first- and second-neighbor positions. For details of the TB model, refer to \cite{fang2015abinitio,zhan2020tunability}.  The interlayer Hamiltonian includes only the interaction between the adjacent chalcogen atoms at the interface of each bilayer, which is  
\begin{align}
\hat{H}^{int}_{12/23} =&  \sum_{\langle p_i{'},\vec{r}_2,p_j,\vec{r}_1\rangle}t_{p_i{'},p_j}(\vec{r}_2-\vec{r}_1)\phi_{2,p_i{'}}^{\dag}(\vec{r}_2)\phi_{1,p_j}(\vec{r}_1) \nonumber \\
&+ H.c.,
\label{equation:interlayer_ham}
\end{align}
where $\phi_{i,p_i}$ is the $p_i$ orbital basis of the $i$th monolayer. The Slater-Koster interlayer hopping term is \cite{fang2015abinitio}
\begin{equation}
t_{p_i{'},p_j}(\vec{r})=[V_{pp,\sigma}(r) - V_{pp,\pi}(r)]\frac{r_i r_j}{r^2}+V_{pp,\pi}(r)\delta_{i,j},
\label{equation:interlayer_hopping}
\end{equation}
where $r=|\vec{r}|=|\vec{r}_2-\vec{r}_1|$, $i, j= x, y, z$, $r_i$, $r_j$ are the components of the relative position vector $\vec{r}$ of orbitals $p_i'$ and $p_j$, respectively, $V_{pp,b}(r)=\nu_b exp[-(r/R_b)^{\eta_b}]$, in which $b$ can be $\sigma$ and $\pi$, and $\nu_b$, $R_b$, and $\eta_b$ are fitting parameters \cite{fang2015abinitio}. The cutoff distance in the Eq. (\ref{equation:interlayer_hopping}) is set to 0.75 nm. In the TB models discussed in Ref. \cite{venkateswarlu2020electronic,vitale2021flat}, interlayer coupling terms beyond the first-neighbor, for example, the $p$ S-$p$ S, $d$ Mo-$p$ S and $d$ Mo-$d$ Mo in the Slater-Koster scheme are considered. The latter two hoppings only have minor changes to the band structure \cite{zhan2020tunability}. Therefore, the main conclusion in this paper will not change if we include also $d$ Mo-$p$ S and $d$ Mo-$d$ Mo interlayer hopping terms in Eq. (\ref{equation:interlayer_ham}) \cite{zhan2020tunability}.  
	
\subsection{Lattice relaxation}
Previous results show that the lattice reconstruction plays a significant role in the electronic structures of twisted bilayer TMDs \cite{naik2020origin,li2021imaging}. To determine the equilibrium structure of a TTM, we perform a lattice relaxation using the LAMMPS software package~\cite{plimpton1995fast, plimpton1995computational}. The potentials we used in relaxation are the interlayer Lennard-Jones (LJ) potential~\cite{jiang2015parametrization} and the intralayer Stillinger-Weber (SW) potential~\cite{rappe1992uff}. During the relaxation process, we use periodic boundary conditions in all three dimensions, leading to both in-plane and out-of-plane displacements. After the relaxation, the interlayer hopping terms are modified via the Slater-Koster scheme in Eq. (\ref{equation:interlayer_hopping}), and intralayer hopping terms are modified with the following formula~\cite{rostami2015theory}:
\begin{equation}
t_{ij,\mu \nu}(\boldsymbol{r}_{ij})=t_{ij,\mu \nu}(\boldsymbol{r}_{ij}^{0})(1-\Lambda_{ij,\mu \nu}\frac{|\boldsymbol{r}_{ij}-\boldsymbol{r}_{ij}^{0}|}{|\boldsymbol{r}_{ij}^{0}|}),
\label{equation:modify}
\end{equation}
where $ t_{ij,\mu \nu}(\boldsymbol{r}_{ij}^{0})$ is the previous intralayer hopping term between the $\mu$ orbital of the $i$ atom and the $\nu$ orbital of the $j$ atom, and  $t_{ij,\mu \nu}(\boldsymbol{r}_{ij})$ is the corresponding new hopping term after relaxation, $\boldsymbol{r}_{ij}^{0}$ and $\boldsymbol{r}_{ij}$ are the relative positions of atoms before and after relaxation, respectively, and $\rm \Lambda_{ij,S-S}=3$, $\rm \Lambda_{ij,S-Mo}=4$, $\rm \Lambda_{ij,Mo-Mo}=5$ stand for $p$ S-$p$ S, $p$ S-$d$ Mo, $d$ Mo-$d$ Mo hybridizations, respectively~\cite{rostami2015theory}.	

\subsection{Electronic properties}
After constructing the Hamiltonian Eq. (\ref{equation:hamiltonian}) of the TTM, we diagonalize the matrix to obtain the band structure and eigenstates, and adopt the tight-binding propagation method to get other electronic structures~\cite{li2023tbplas}. 
The DOS can be calculated by \cite{yuan2010modeling, li2023tbplas}
\begin{align}
D(\mathcal{E})=\lim_{S \to \infty}\frac{1}{S}\sum_{p=1}^S \frac{1}{2\pi}\int_{-\infty}^{\infty}e^{i \mathcal{E} t} \left\langle \psi_p | e^{-iHt} | \psi_p \right\rangle dt,
\end{align}
where $\left|\psi_p \right\rangle$ is the random initial state and S is the number of random samples. To guarantee the convergence of DOS, we calculate the DOS of a large system with more than ten millions of orbitals.
 
\section{Results and discussion}
\label{sec3}	
\subsection{The interlayer interaction effect}
\label{band}
\begin{figure*}[htbp]
    \centering
    \includegraphics[width=\textwidth]{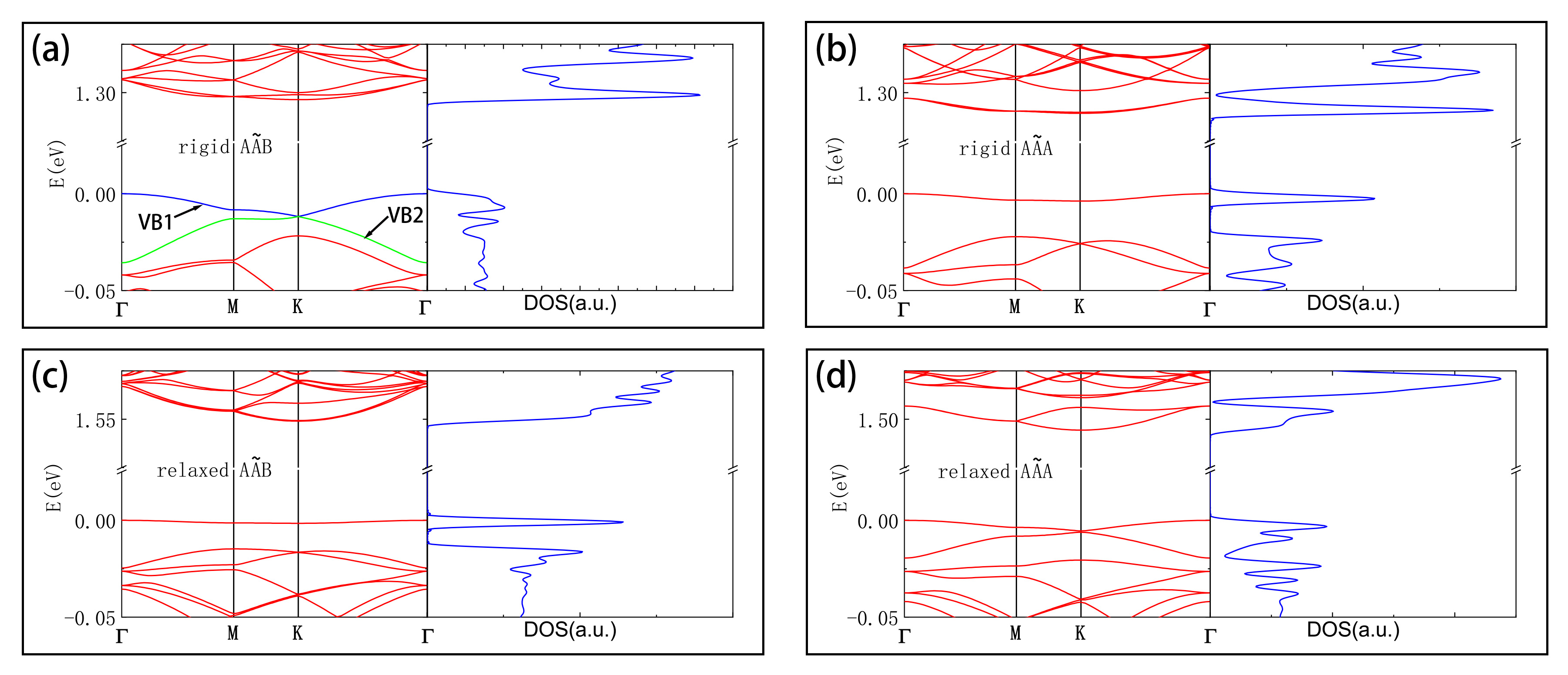}
    \caption{The band structures and DOS of $\rm A\Tilde{A}B-3.15^\circ$ and $\rm A\Tilde{A}A-3.15^\circ$. (a) and (c) are band structures and DOS of rigid and relaxed $\rm A\Tilde{A}B$, respectively. (b) and (d) are band structures and DOS of rigid and relaxed $\rm A\Tilde{A}A$, respectively.}
    \label{fig:bands}
\end{figure*}
\begin{figure*}[htbp]
    \centering
    \includegraphics[width=\textwidth]{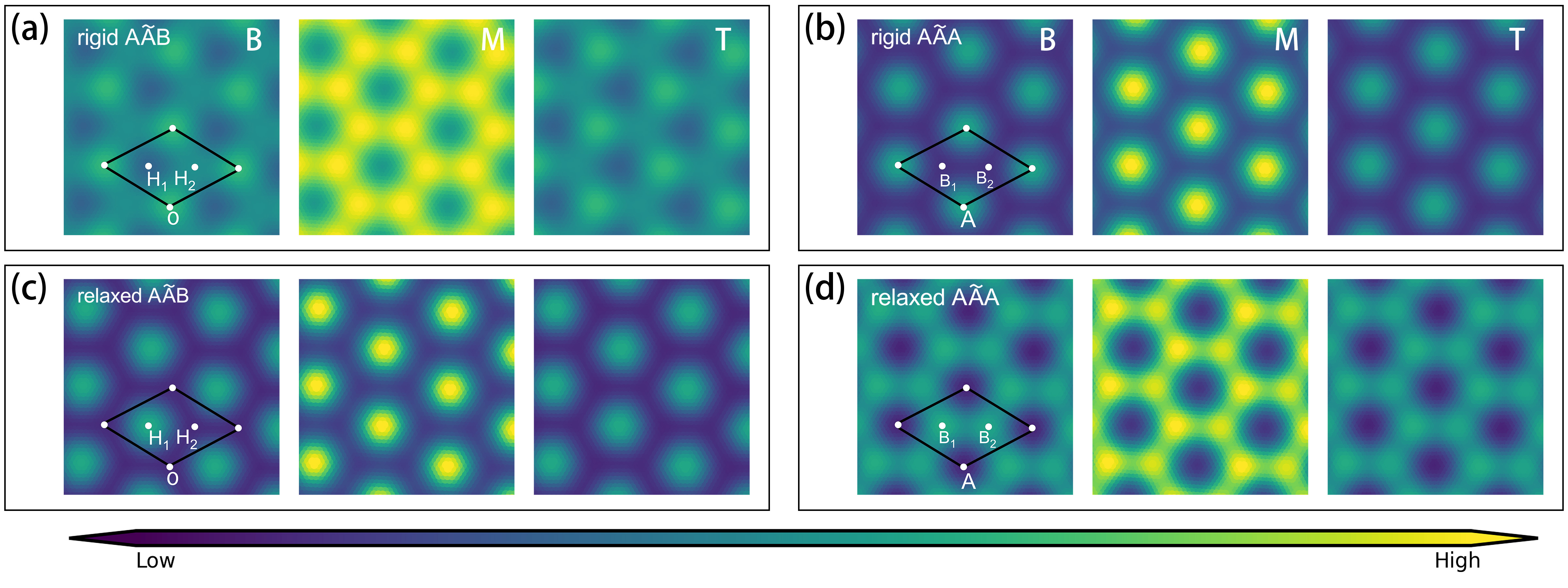}
    \caption{The eigenstates of the flat band at the $\Gamma$-point in $\rm A\Tilde{A}B-3.15^\circ$ and $\rm A\Tilde{A}A-3.15^\circ$. (a) and (c) are eigenstates of rigid and relaxed $\rm A\Tilde{A}B$, and (b) and (d) are eigenstates of rigid and relaxed $\rm A\Tilde{A}A$, respectively. In the first image on the left side of each rectangular black box, the unit cell and high-symmetry points are marked with black diamond frames and white dots, respectively. In each black rectangle, the panels from left to right are states in the bottom (B), middle (M) and top (T) layers, respectively.}
    \label{fig:eigenstates}
\end{figure*}
The electronic structures of the tb-TMDs have been systematically investigated \cite{naik2018ultraflatbands,zhan2020tunability,vitale2021flat,venkateswarlu2020electronic}. In tb-TMDs with a AA arrangement, the high-symmetry stackings include AA regions and two types of Bernal-like regions which are $\mathrm{B^{M/X}}$ and $\mathrm{B^{X/M}}$. In the rigid AA case, the first (VB1) and the second (VB2) highest valence bands (VBs) are separated by a gap at the K-point of BZ. 
We refer to this gap as $\Delta_2$, defined as the energy difference between the minimum of VB1 at the K-point and the maximum of VB2 at either the K- or M-point. The state of the VB1 at the $\Gamma$-point is localized in the AA region. When the lattice relaxation is taken into account, the VB1 and VB2 touch at the K-point and the states of the VB1 at the $\Gamma$-point localize in the $\mathrm{B^{M/X}}$ and $\mathrm{B^{X/M}}$ regions. The situation of the tb-TMDs with a AB arrangement is completely different. In the tb-TMDs with a AB structure, the high-symmetry stackings are AB, $\mathrm{B^{X/X}}$ and $\mathrm{B^{M/M}}$. In both rigid and relaxed cases, there is a gap $\Delta_2$, with the states at the $\Gamma$-point in the $\mathrm{B^{X/X}}$ and AB regions, respectively. Moreover, in both AA and AB cases, the moir\'e systems exhibit a semiconducting band structure with a band gap separating the valence and conduction bands. We define this gap as $\Delta_1$, which is the energy difference between the $\Gamma$-point of the VB1 and the K-point of the conduction band edge. Note that we focus on the valence band structure of the twisted trilayer TMDs. For simplicity, we will not include the SOC in the calculations unless we discuss about the SOC effect.  

As for the homotrilayer systems, the $\rm A\Tilde{A}B$ is a combination of AA and AB moir\'e patterns, and the $\rm A\Tilde{A}A$ is a combination of two AA moir\'e patterns, which are justified by the electronic structures in Figs. \ref{fig:bands}(a), (b) and Figs. \ref{fig:eigenstates}(a), (b). In the $\rm A\Tilde{A}B-3.15^\circ$, the VB1 and VB2 touch at the $K-$point, with the states at the $\Gamma$-point localized in both O (corresponding to the AA bilayer case) and H$_2$ (corresponding to the AB bilayer case) regions. In $\rm A\Tilde{A}A-3.15^\circ$, similar to the AA bilayer case, the VB1 is isolated from other VBs, and the states at the $\Gamma$-point of the VB1 are in the A region [illustrated in Fig. \ref{fig:structure}(d)] with an AAA stacking. The AAB arrangement could be constructed by a lateral shift of the AAA arrangement along the armchair direction. Therefore, the lateral shift is an efficient tuning node in the TTM, for instance, opening a gap between the VB1 and VB2, engineering the localization and the width of the flat bands.   

The interlayer interaction could be modified by a lateral shift. Another natural way is lattice relaxation. The out-of-plane displacements of the moir\'e patterns lead to local variations in the interlayer spacing (ILS) between two layers in different high-symmetry regions, resulting in the variation of the interlayer interaction. In the tb-TMDs, the lattice relaxation has a significantly effect on the flat bands \cite{naik2018ultraflatbands,zhan2020tunability,vitale2021flat}. Next, we study how the lattice relaxation will modify the flat band properties of TTM. Our findings reveal that in the relaxed $\rm A\Tilde{A}B$, the VB1 becomes an isolated flat band, and states are localized in the H$_1$ with S-Mo-Mo stacking. On the contrary, in the relaxed $\rm A\Tilde{A}A$ structure, the VB1 and VB2 touch at the $K-$point, forming a Dirac-like bands. The states of the flat band at the $\Gamma$-point are mainly distributed in the B$_1$ and B$_2$ high-symmetry points. Therefore, in the band structures plotted in Fig. \ref{fig:bands}, there are two different types of flat bands. One is the isolated flat bands, of which the states are mainly localized in one symmetry-stacking point; the other flat band touches with VB2 at the $K-$point, forming a Dirac-like dispersion, and states at different $k$ points have different localization (see Appendix \ref{eigenstates_appendix}). 
Moreover, we find that the presence or absence of the gap $\Delta_2$ does not change if different potentials \cite{naik2019kolmogorov} are used in the lattice relaxation process and the results are both consistent with the DFT results in Appendix \ref{dft_results}.

\begin{figure}[htbp]
    \centering
    \includegraphics[width=0.5\textwidth]{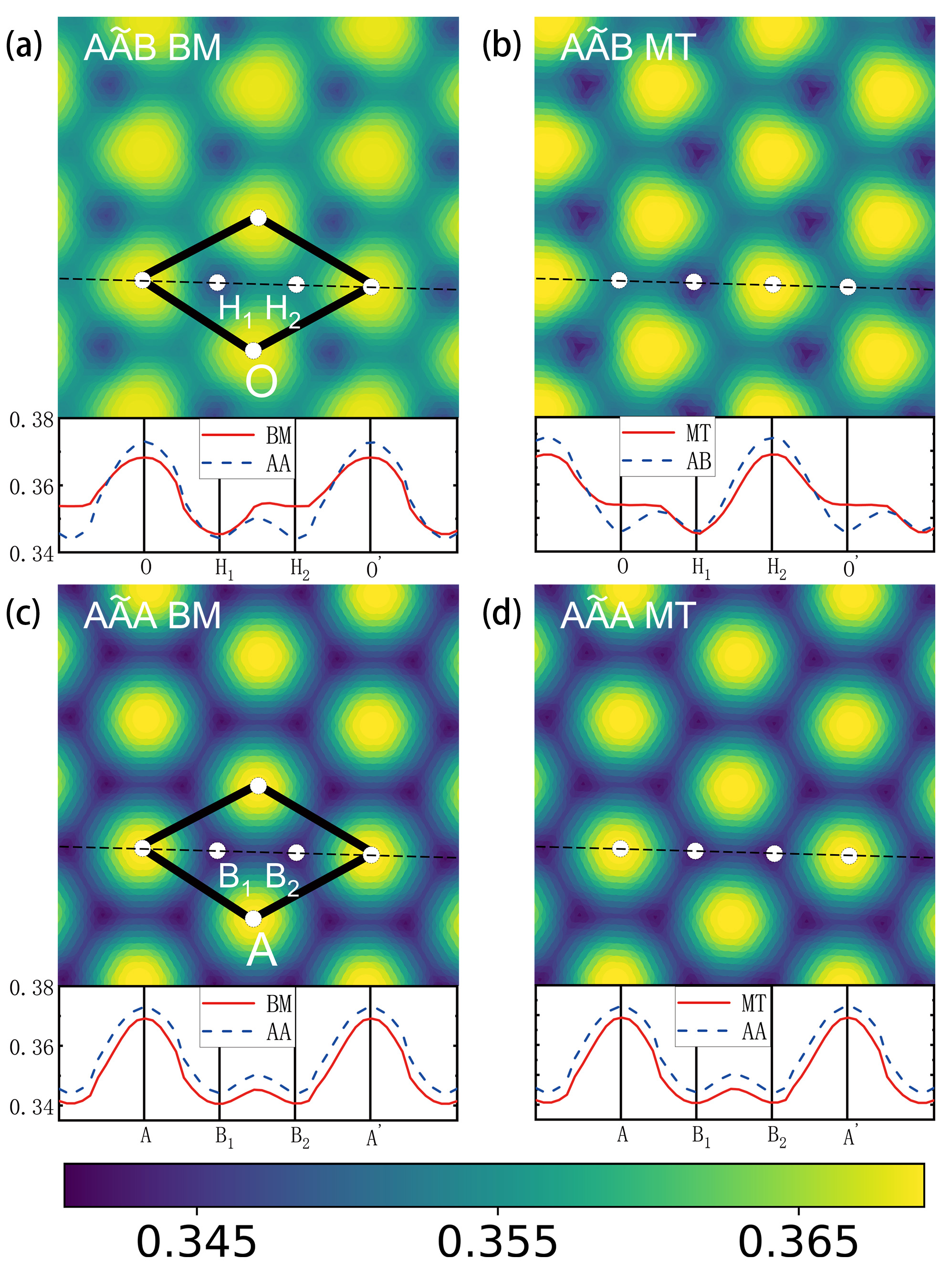}
    \caption{The interlayer separations (ILSs) of $\rm A\Tilde{A}B$ and $\rm A\Tilde{A}A$. Parts (a) and (b) are ILSs of relaxed $\rm A\Tilde{A}B$, and (c) and (d) are ILSs of relaxed $\rm A\Tilde{A}A$. Parts (a) and (c) are the ILS between the bottom and middle layers, and (b) and (d) are the ILS between the middle and top layers. Here, ILS is defined as the distance in the out-of-plane direction between adjacent S layers.}
    \label{fig:ILS}
\end{figure}

To better understand the formation of the flat bands, we analyze the ILSs of the moir\'e systems, which are shown in Fig. \ref{fig:ILS}. For comparison, we also plot the ILSs of the tb-TMDs with AA and AB configurations (dashed lines).  In the relaxed $\rm A\Tilde{A}B$, the maximum ILS between the bottom layer and middle layer is at O with AA stacking, while the maximum ILS between the middle layer and top layer is at H$_2$ with S-S stacking. Due to the presence of the third layer, the ILSs of the H$_2$ and O regions are different from that of the bilayer cases. In particular, the H$_2$ and O in Figs. \ref{fig:ILS}(a) and \ref{fig:ILS}(b) have larger ILSs, respectively. On the contrary, the third layer has no effect on the ILS in the H$_1$ regions. Therefore, there is a competition between the AA and AB bilayer patterns in the lattice reconstruction. The homotrilayer AAA exhibits similar relaxation patterns to homobilayer AA. That is, there is only a global reduction of the ILS in the whole moir\'e pattern.   

\begin{figure*}[htbp]
    \centering
    \includegraphics[width=\textwidth]{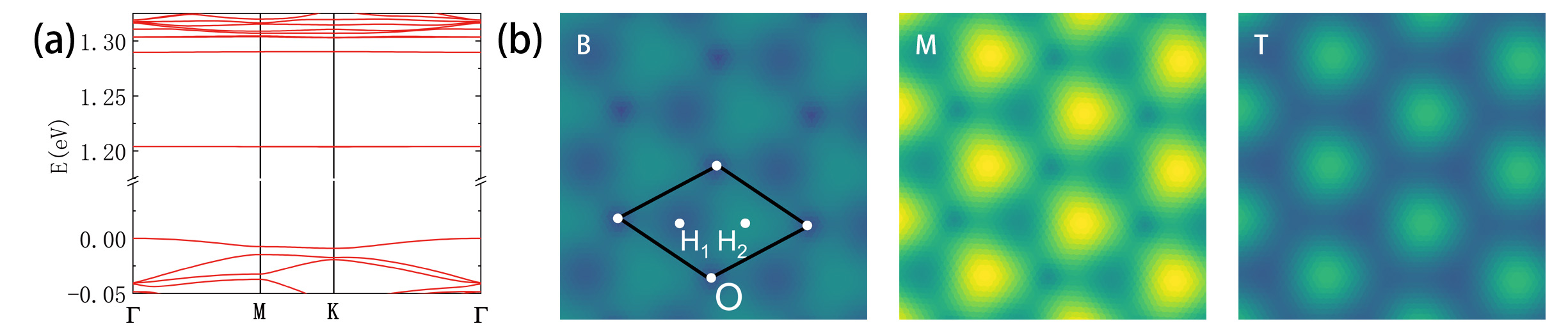}
    \caption{The band structures and eigenstates of rigid $\rm A\Tilde{A}B$ with modified ILS. Parts (a) and (b) are the band structure and eigenstates of rigid $\rm A\Tilde{A}B$ with a modified ILS of AA stacking in the O region, respectively. Here, the ILS is modified with a Gauss bubble, and the amplitude of the Gauss bubble is $h = 0.05\; c$, the center of the Gauss bubble is O, and the width of the Gauss bubble is $\rm (\sigma_x, \rm \sigma_y) = (0.5, 0.5)$ nm. The color maps in (b) is the same as that in Fig. \ref{fig:eigenstates}.}
    \label{fig:bubble}
\end{figure*}

The change in electron distribution arises from the changes of the interlayer interaction in different high-symmetry stacking regions. It will be conformed by the strain effect in the following section. In the rigid form of $\rm A\Tilde{A}B$, the strongest interlayer interaction between the bottom and middle layers occurs in the O point with AA stacking, and the strongest interlayer interaction between the middle and top layers occurs in the H$_2$ point with S-S stacking. Different from the AAA stacking in rigid $\rm A\Tilde{A}A$, there are two dominant stacking forms in rigid $\rm A\Tilde{A}B$. As a result, the electron distribution of the bottom layer is concentrated in the O point, the electron distribution of the top layer is concentrated in the H$_2$ point, and the electron distribution of the middle layer is present at both O and H$_2$ points, as depicted in Fig. \ref{fig:eigenstates}(a). Hence, the electrons do not exhibit localization in the TTM plane, and VB1 represents a non-isolated band in the band structure of rigid $\rm A\Tilde{A}B$ in Fig. \ref{fig:bands}(a). After relaxation, electrons are redistributed to the H$_1$ point, where interlayer interactions are relatively stronger, resulting in the localization of VB1 in relaxed $\rm A\Tilde{A}B$ in Fig. \ref{fig:bands}(c). Similarly, for the $\rm A\Tilde{A}A$ structure, in its rigid form, the strongest interlayer interaction between the bottom and middle layers, as well as between the middle and top layers, occurs in the O point. Consequently, electrons in all three MoS$_2$ layers are localized in the O point, and VB1 of rigid $\rm A\Tilde{A}A$ represents an isolated band in Fig. \ref{fig:bands}(b). After relaxation, the interlayer interaction in the O point weakens, and electrons are redistributed to the H$_1$ and H$_2$ points, causing a transition in VB1 from an isolated to a non-isolated band in Fig. \ref{fig:bands}(d).
 
\subsection{The strain effect}
\label{deformation}
	
We use a Gaussian bubble to modify the interlayer separation in the high-symmetry stacking of TTM, increasing the interlayer orbital distance of the AA stacking in the O point or the S-S stacking in the H$_2$ point, respectively. The Gaussian function is in the form of $\rm \Delta ^{ILS} = \mathit h e^{-(x-x_0)^2/\sigma_x^2-(y-y_0)^2/\sigma_y^2}$, where $h$ represents the amplitude of the bubble, $\rm (x_0, y_0)$ denotes the center of the Gauss bubble, and $\rm \sigma_x$ and $\rm \sigma_y$ are the widths of the bubble in the x and y directions. The center of the Gaussian function is at either O or H$_2$ (calculated separately for the two cases). For example, when the center is at O, all orbitals on the bottom layer inside the bubble have their $z-$coordinates subtracted by $\rm \Delta ^{ILS}$, causing a movement away from the middle layer and top layer, while the middle and top layers remain unchanged. For the hopping terms, we still calculate them as we do during relaxation, with interlayer hopping terms calculated according to the Slater-Koster formula in Eq. (\ref{equation:interlayer_hopping}) and intralayer hopping corrected using Eq. (\ref{equation:modify}). 

We analyze the electronic properties of rigid $\rm A\Tilde{A}B$ with the modified interlayer separation. At the high-symmetry stacking points O and H$_2$ in rigid $\rm A\Tilde{A}B$, the stacking forms are AAB and Mo-S-S, respectively. By increasing the interlayer separation and weakening the interlayer interaction between orbitals in AA stackings at O or S-S stacking at H$_2$, we calculate the energy bands and eigenstates of $\rm A\Tilde{A}B$, shown in Fig. \ref{fig:bubble}. The corresponding band structure and eigenstates for weakened interlayer interaction in S-S stacking are not shown here as they exhibit similar behavior to Fig. \ref{fig:bubble}. When the interlayer interaction in AA stacking is weakened, electrons become localized at H$_2$. Conversely, and when the interlayer interaction in S-S stacking is weakened, electrons are localized at O. Both of these operations cause electrons to exhibit localization in the TTM plane, and the VB1 of $\rm A\Tilde{A}B$ becomes an isolated energy band. This calculation supports our hypothesis that electrons primarily distribute at high-symmetry points with strong interlayer interactions. If there is only one high-symmetry point with dominant interlayer interaction, VB1 will be an isolated band, while if multiple high-symmetry points contribute, VB1 will touch with VB2 at the $K-$point.

\subsection{The spin-orbital coupling effect}
\label{SOC}
\begin{figure}[htbp]
    \centering
    \includegraphics[width=0.5\textwidth]{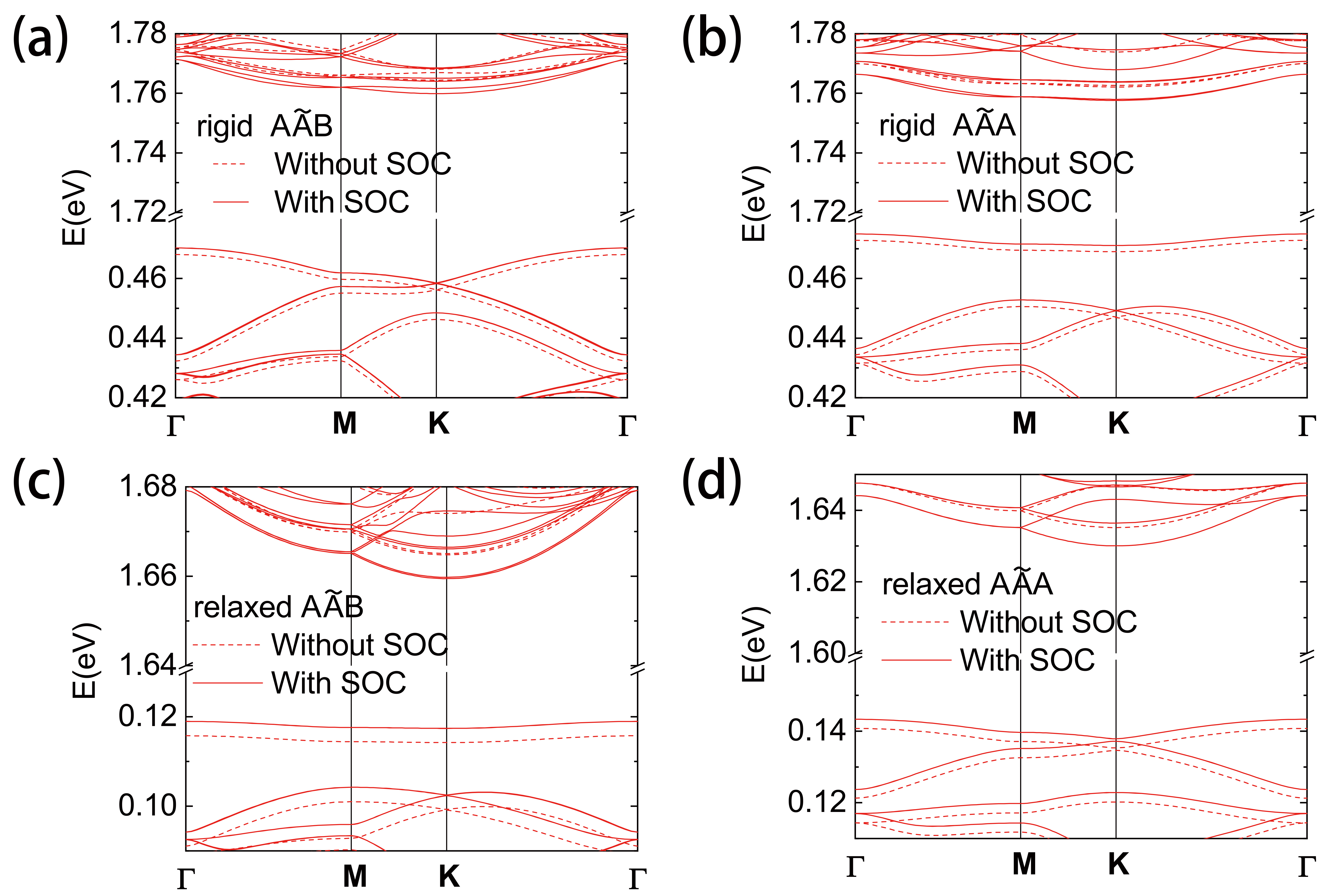}
    \caption{The band structures with and without SOC of $\rm A\Tilde{A}B$ and $\rm A\Tilde{A}A$ with and without relaxation. Parts (a) and (c) are band structures of rigid and relaxed $\rm A\Tilde{A}B$, respectively, and (b) and (d) are band structures of rigid and relaxed $\rm A\Tilde{A}A$, respectively.
    }
\label{fig:bands_soc}
\end{figure}

\begin{figure*}[htbp]
    \centering
    \includegraphics[width=\textwidth]{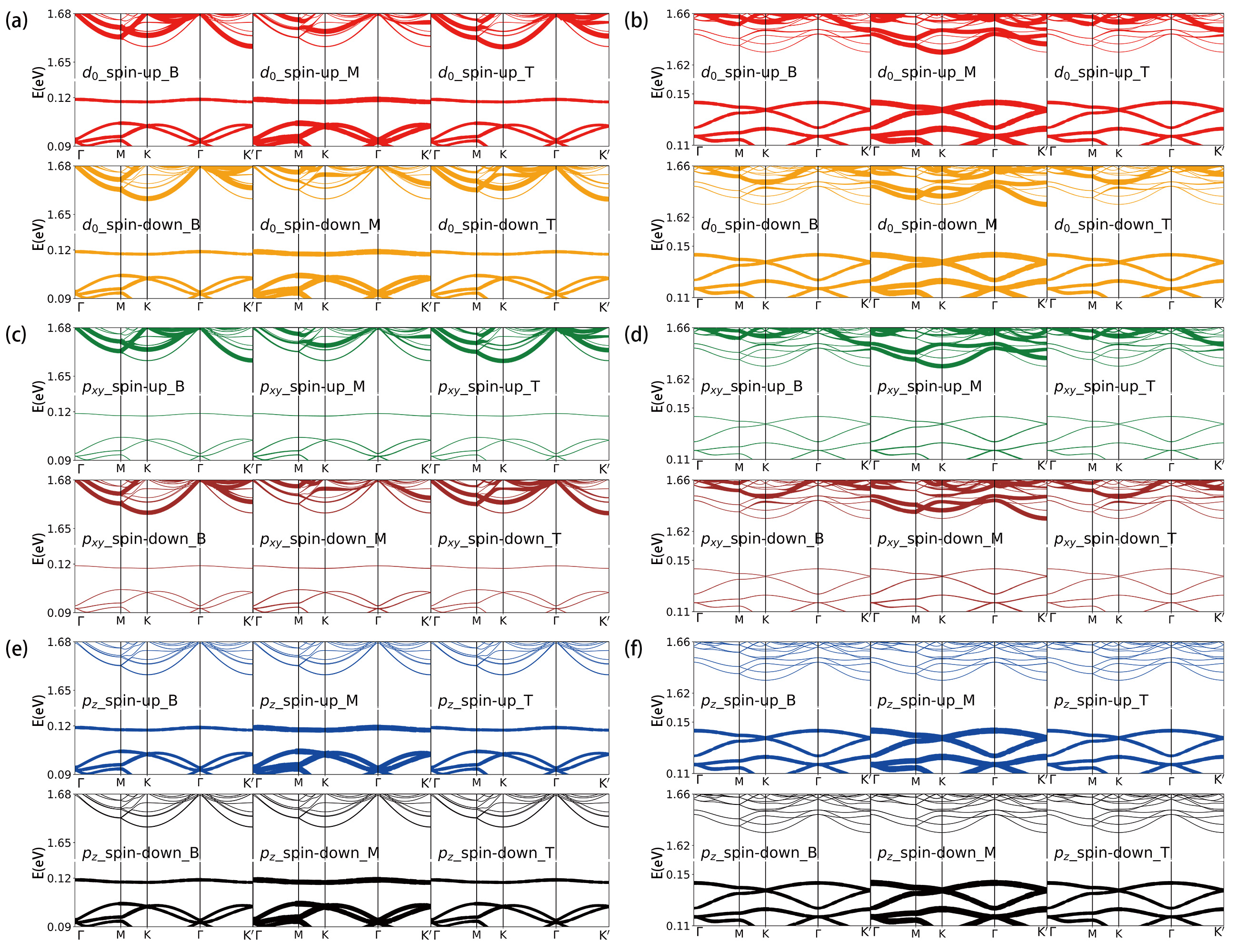}
    \caption{The orbital weights of relaxed $\rm A\Tilde{A}B$ and relaxed $\rm A\Tilde{A}A$. The $d$ character ($d_0=d_{z^2}$, $d_1=d_{xz},d_{yz}$, $d_2=d_{x^2-y^2}, d_{xy}$) refer to the 5 $d$ orbital of Mo atoms, and $p$ ($p_{xy}=p_x + p_y$, $p_z$) refers to the $p$ orbital of S atoms. Parts (a), (c), and (e) are the orbital weights of  spin-up and spin-down d$_0$, p$_{xy}$, and p$_z$ orbitals in relaxed $\rm A\Tilde{A}B$. Parts (b), (d), and (f) are the orbital weights of  spin-up and spin-down d$_0$, p$_{xy}$, and p$_z$ orbitals in relaxed $\rm A\Tilde{A}A$. The orbital weight values of $d_2$ and $d_1$ are not shown here since they have very few contributions to the band edges.}
    \label{fig:weights}
\end{figure*}
	
In this section, we investigate the influence of the SOC effect on the band structures of TTM. We introduce the SOC in the TB Hamiltonian by doubling the orbitals and adding an atomic on-site term $\rm \lambda_{SO}^{Mo/S} \bm{L}\cdot \bm{S}$ \cite{fang2015abinitio, zhan2020tunability, roldan2014momentum}. In Fig. \ref{fig:bands_soc}, we show the band structures of $\rm A\Tilde{A}B$ and $\rm A\Tilde{A}A$ with SOC (solid red line) and without SOC (dashed red line). It is evident that regardless of whether the structure is rigid or relaxed, the valence-band edge does not split due to SOC, which means the band at the valence-band edge is doubly spin degenerate in the BZ. This finding is contrary to that of monolayer MoS$_2$ and consistent with tb-TMDs \cite{fang2015abinitio,zhan2020tunability}.  In contrast, there is a large splitting of the energy band at the $K-$point of the conduction band edge, and at the $\Gamma$-point and $M-$point with time reversal symmetry, the energy bands remain spin degenerate. Moreover, the SOC decreases the $\Delta_1$ gap in all cases due to the splitting of the conduction band.

To understand the SOC effect on the flat bands, we analyze their orbital weights. Figure \ref{fig:weights} presents the orbital weights of relaxed $\rm A\Tilde{A}B$ and relaxed $\rm A\Tilde{A}A$. We use the thickness of the lines to represent the size of the weight, and draw separately the weight diagrams for spin-up and spin-down, as well as the orbital weights of the three layers. It is obvious that the d$_{z^2}$ and p$_z$ orbitals play a major role in the formation of the ultraflat bands at the valence-band edge, and the contribution of d$_{z^2}$ and p$_z$ orbitals in the middle layer is greater than that in the bottom and top layers. This is consistent with the results in Fig. \ref{fig:eigenstates}, which indicates that the electron distribution in the middle layer is the result of the combined action of the interlayer interaction between the bottom and middle layers and that between the top and middle layers. In monolayer TMD, the valence-band edge at the $K-$point has mainly $d_2$ orbital character, whereas at the $\Gamma$-point it has mainly $d_0$ and $p_z$ orbital characters \cite{cappelluti2013tight}. This suggests that the flat band in $\rm A\Tilde{A}B$ originate from the $\Gamma$-states of the constituent monolayer. All these behaviors of the flat band at the valence-band edge are also detected in the $\rm A\Tilde{A}A$ case.

\begin{figure*}[htbp]
    \centering
    \includegraphics[width=\textwidth]{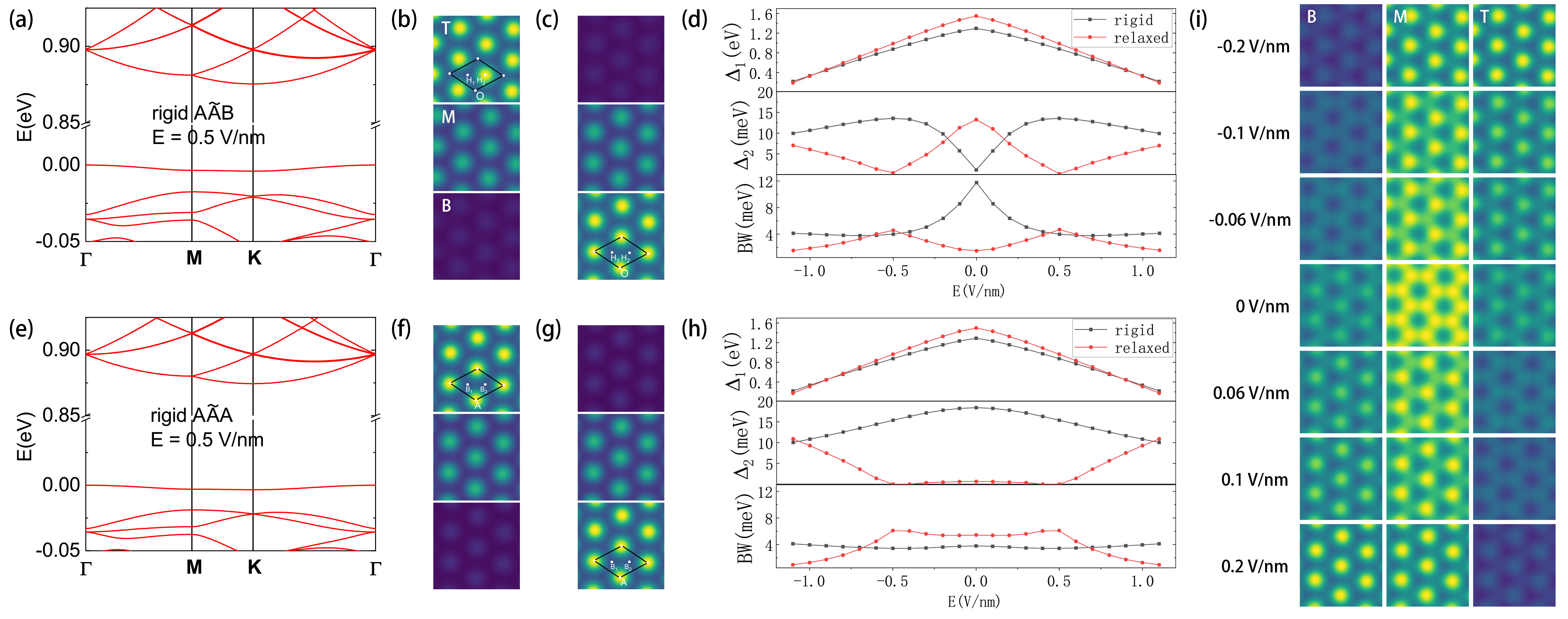}
    \caption{The electric field response. The band structure, eigenstates, and flat band response (black squares) for (a)-(d) the rigid $\rm A\Tilde{A}B$ case, and (e)-(h) the rigid $\rm A\Tilde{A}A$ case. The cases for relaxed $\rm A\Tilde{A}B$ and $\rm A\Tilde{A}A$ (red dots) are also plotted in (d) and (h), respectively. The figures in (i) are the eigenstates at the $\Gamma$-point of rigid $\rm A\Tilde{A}B$ with the electric field -0.2, -0.1, -0.06, 0, 0.06, 0.1 and 0.2 V/nm, from the left to the right are the eigenstates in the bottom, middle, and top layers. The electric field applied in (a), (c), (e) and (g) is 0.5 V/nm while the electric field used in (b) and (f) is -0.5 V/nm. In (b), (c), (f) and (g), the figures from the bottom to the top are eigenstates at the $\Gamma$-point in bottom, middle and top layers, respectively. 
    The band structure under an electric field of magnitude -0.5 V/nm is similar to that under an electric field of magnitude 0.5 V/nm, so it is not included in this article. The color maps in (b), (c), (f), (g) and (i) use the same colorbar as Fig. \ref{fig:eigenstates}}
    \label{fig:electric}
\end{figure*}     

The bottom of the conduction band mainly consists of $\rm d_0$ and $\rm p_{xy}$ orbitals, and from Figs. \ref{fig:weights}(a) and \ref{fig:weights}(c), it is found that at the bottom of the conduction band of relaxed $\rm A\Tilde{A}B$, the spin-up state at the $K-$valley is degenerate with the spin-down state at the $K^\prime$-valley, and vice versa, indicating that the spins are locked to a certain valley, which is a spin-valley locking effect \cite{zhan2020tunability, schneider2019spin}. Spin-valley locking also occurs in the electronic structure of relaxed $\rm A\Tilde{A}A$ shown in Figs. \ref{fig:weights}(b) and \ref{fig:weights}(d). The spin-valley locking effects in relaxed $\rm A\Tilde{A}B$ and relaxed $\rm A\Tilde{A}A$ are similar to the case of tb-MoS$_2$ \cite{zhan2020tunability}. 

The three-layer MoS$_2$ orbitals have a great difference in the contribution to the conduction band edge. The middle layer contribute the most in relaxed $\rm A\Tilde{A}A$, while the bottom layer or top layer contribute the most to the bottom of the conduction band in relaxed $\rm A\Tilde{A}B$, which means that the electrons in $K-$ or $K^\prime$-valley are localized in one layer of TTM. This is because the interlayer hopping is suppressed due to the spin-valley locking \cite{xu2014spin, zhang2023every, bg2020spin}. The influence of interlayer interaction on the electron distribution is greatly reduced, and electrons are localized in a certain layer, depending on the spin-valley state and layer index. 

In relaxed $\rm A\Tilde{A}B$ the spin-up states in the bottom layer and spin-down states in the top layer are degenerate, and vice versa, indicating that there is a spin-layer locking in relaxed $\rm A\Tilde{A}B$. This is because in the $\rm A\Tilde{A}B$ structure, the middle layer is twisted, and the bottom and top layers form a bilayer structure with 2H phase, which has inversion symmetry in real space. The inversion symmetry brings the spin-layer locking, which is similar to the results of the electronic band structure of bilayer TMDs \cite{schneider2019spin, xu2014spin, jones2014spin, xu2021ab, bg2020spin}. Most of electrons in relaxed $\rm A\Tilde{A}B$ are locked in a certain layer and valley, which is the so called spin-valley-layer locking \cite{xu2021ab, jiang2017zeeman}, and can be described by the coupling between electron real spin, valley pseudospin  and layer pseudospin. The Hamiltonian near the $K-$ and $K^\prime$-point can be written as
$\hat{H}=-\lambda \tau_z \hat{S}_z \hat{\sigma}_z+t_{\perp} \hat{\sigma}_x$  \cite{xu2014spin, jiang2017zeeman, gong2013magnetoelectric}, 
in which $\lambda$ is the SOC splitting amplitude, $\tau_z=\pm 1$ is the index of the $K-$ and $K^\prime$-valley pseudospin, $\hat{S}_z=\pm 1$ is the index of spin up and down, $t_{\perp}$ is the interlayer hopping amplitude, and $\hat{\sigma}$ are the Pauli matrices for the layer pseudospin. We set the $\hat{\sigma}_z$ of layer A to -1, the $\hat{\sigma}_z$ of layer B to 1, and the $\hat{\sigma}_z$ of the twisted middle layer A to a value $\alpha$ between 0 and 1 related to the twist angle. In the band structure of relaxed $\rm A\Tilde{A}B$, the energy band at the bottom of the conduction band splits into two energy bands at $K-$ and $K^\prime$-valley due to the SOC effect. For the energy band with larger values, $\tau_z\hat{S}_z\hat{\sigma}_z=-1$ is satisfied, while for the energy band with smaller values, $\tau_z\hat{S}_z\hat{\sigma}_z=1$ is satisfied. Thus, a spin in a valley is locked to a certain layer, indicating a spin-valley-layer locking in the bottom layer and top layer of relaxed $\rm A\Tilde{A}B$. Meanwhile, in the band structure of relaxed $\rm A\Tilde{A}A$, there is also two splitting bands at the bottom of conduction band. For the energy band with larger values, $\tau_z\hat{S}_z=-1$ is satisfied, while for the energy band with smaller values, $\tau_z\hat{S}_z=1$ is satisfied, and $\hat{\sigma}_z=\alpha$, which means there is a spin-valley locking in relaxed $\rm A\Tilde{A}A$, and there is no spin-layer locking because of the lack of inversion symmetry in $\rm A\Tilde{A}A$. The spin-valley-layer locking effect makes $\rm A\Tilde{A}B$ a promising platform for studying optical and magnetoelectric effects involving the degrees of freedom including the real electron spin, the layer pseudospin, and the valley pseudospin \cite{khani2020gate, zhang2023every, gong2013magnetoelectric, xu2014spin, jones2014spin, bg2020spin}.

\begin{figure*}[htbp]
    \centering    
    \includegraphics[width=\textwidth]{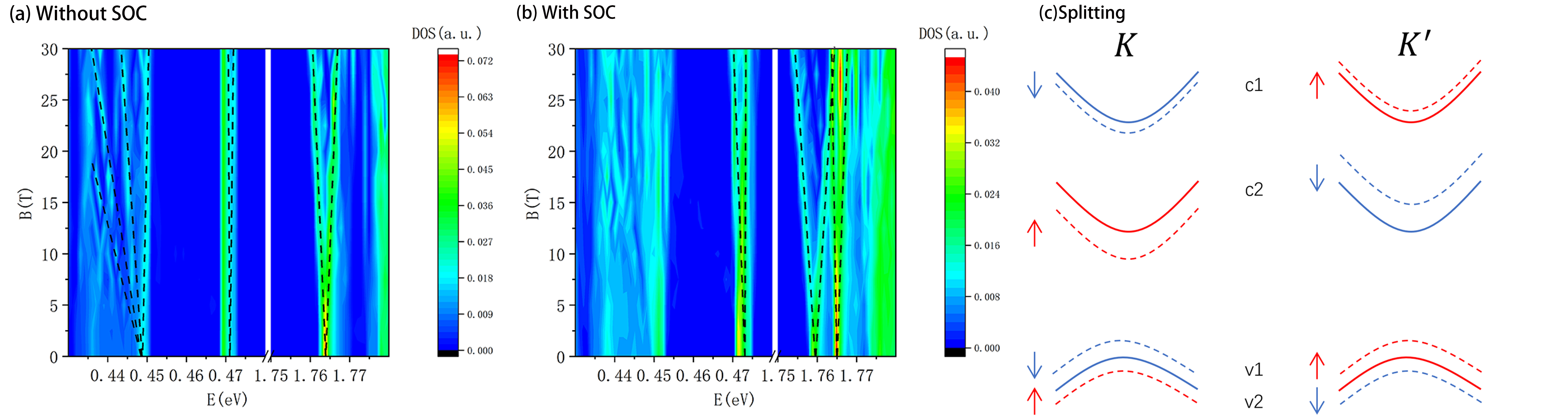}
    \caption{The magnetic field effect on the rigid $\rm A\Tilde{A}A$. Parts (a) and (b) are the DOS versus energy and magnetic fields without and with SOC, respectively. In (b), we also consider the spin Zeeman effect. (c) The schematic diagram of the energy level splitting of rigid $\rm A\Tilde{A}A$ (solid curves) with SOC under a magnetic field. The dashed curves represent the Zeeman-shifted bands. We ignore the Landau Levels in (b).}  
    \label{fig:dos_magnetic}
\end{figure*}
\subsection{The electric field effect}
\label{electic_field}

In this section, we study the response of electronic properties of TTM to the electric field. We define the electric field as positive when its direction is from the bottom to the top of TTM, and negative when it is from the top to the bottom. The effect of the electric field is considered by adding an electric potential on the onsite term in Eq. (\ref{equation:hamiltonian}). 
As shown in Figs. \ref{fig:electric}(a) and (e), the VB1 of rigid $\rm A\Tilde{A}B$ and rigid $\rm A\Tilde{A}A$ become an isolated band when the electric field is 0.5 V/nm  or -0.5 V/nm (the band structure for an electric field of -0.5 V/nm is not shown here). For rigid $\rm A\Tilde{A}B$, when the electric field is -0.5 V/nm, the electrons are localized at H$_2$, and when the electric field is 0.5 V/nm, the electrons are localized at O, as shown in Figs. \ref{fig:electric} (b) and \ref{fig:electric} (c). In contrast, Figs. \ref{fig:electric} (e) and \ref{fig:electric} (f) show that regardless of whether the electric field is -0.5 V/nm or 0.5 V/nm, the electrons are localized at A. 


We investigate how the gaps $\Delta_1$, $\Delta_2$ (defined in Sec. \ref{band}), and the bandwidth of VB1 (BW) change with the electric field, as depicted in Figs. \ref{fig:electric} (d) and \ref{fig:electric} (h). The $\Delta_1$ of both rigid $\rm A\Tilde{A}B$ and rigid $\rm A\Tilde{A}A$ decrease linearly as the electric field increases, and this tendency occurs regardless of whether the electric field is positive or negative. With the electric field large enough, there will be a transition between the semiconductor and metal. Such giant Stark effect is similar to the case of tb-TMDs \cite{khoo2004tuning,zhang2020tuning}. We can find that the $\Delta_2$ of rigid $\rm A\Tilde{A}B$ increases first and then decreases as the electric field increases, and reaches its maximum value around 0.5 V/nm, while BW has the opposite trend to $\Delta_2$, reaching its minimum value around 0.5 V/nm. In addition, the $\Delta_2$ of rigid $\rm A\Tilde{A}A$ gradually decreases as the electric field increases, while the BW of rigid $\rm A\Tilde{A}A$ changes very little. Furthermore, we observe that $\Delta_1$, $\Delta_2$ and BW are all symmetrical about E=0. In the rigid $\rm A\Tilde{A}A$, the electric field attempts to close the gap between the VB1 and VB2. In order to understand the nonlinear changes of the $\Delta_2$ with the electric field in the $\rm A\Tilde{A}B$ case, we calculate the flat band states in real space, shown in Fig. \ref{fig:electric}(i). There is a charge transfer between the three layers. A positive electric field deplete the top layer, and the system behaves like a tb-TMDs with AA arrangement. That is why the flat band states localize in the O region. Under a negative electric field, the system behaves like a tb-TMDs with AB arrangement, where the flat band states localize in the H$_2$ region. The critical electric field is around 0.5 V/nm. In Figs. \ref{fig:electric}(d) and \ref{fig:electric}(h), we investigated the response of the electronic properties of relaxed $\rm A\Tilde{A}B$ and $\rm A\Tilde{A}A$ to electric fields. We observed a turning point in the behavior of relaxed $\rm A\Tilde{A}B$ and $\rm A\Tilde{A}A$ near $\rm E=\pm0.5 V/nm$. The lattice relaxation leads to increased interlayer separation in regions of originally strong interlayer interaction, resulting in greater on-site energy when an electric field is applied. As the electric field increases, electrons transfer from areas of strong interlayer interaction to those with larger interlayer spacing due to the electric field effect. \textcolor{red}{Moreover, there is a competition between the interlayer interaction and the electric field, resulting in a $W$-shape and "screening effect" of the $\Delta_2$ in relaxed $\rm A\Tilde{A}B$ and $\rm A\Tilde{A}A$, respectively. In the relaxed $\rm A\Tilde{A}B$, the interlayer distance in the O is larger than that of the H$_1$. So the O region has a larger response to the electric field. Consequently, around $\rm E=\pm0.5 V/nm$, the states are localized at both O and H$_1$ regions (not shown here).  The relaxed $\rm A\Tilde{A}A$ structure has similar behavior. After the turning point $\rm E=\pm0.5 V/nm$, there is a charge transfer from points B$_1$ and B$_2$ to the A region where the electric field effect is stronger.}
\subsection{The magnetic field effect}
\label{magnetic_field}

We apply a magnetic field perpendicular to the plane of the TTM via Peierls substitution \cite{graf1995electromagnetic, matsuura2016theory}.
Figure \ref{fig:dos_magnetic}(a) reveals energy level splitting in both the conduction and valence bands of rigid $\rm A\Tilde{A}A$ in the absence of SOC. The valence-band edge has $d_{z^2}$ and $p_z$ orbital characters, which have zero magnetic quantum number. The equal-energy splittings in the valence-band edge are Landau levels. On the contrary, 
there is a valley splitting in the conduction-band edge (illustrated by dashed lines). The valley splitting magnitude is $\mu_e\approx0.105$ meV/T, and the $g-$factor is $g_c\approx\mu_e/\mu_B=1.814$, where $\mu_B=0.05788$ meV/T represents the Bohr magneton. Therefore, the conduction band edge that consists of $d_{z^2}$ and $p_{xy}$ orbitals have both Landau Levels and valley splitting.

When consider the SOC, we also include a spin Zeeman term \cite{xuan2020valley} in the Halmitonian of Eq. (\ref{equation:hamiltonian}). As discussed in Sec. \ref{SOC}, the SOC effect results in a spin splitting of the conduction band edge. 
In the presence of the magnetic field, as shown in Fig. \ref{fig:dos_magnetic}(b), we observe that the SOC-induced energy levels at the bottom of the conduction band not only manifests as valley splitting, but they also experiences a shift attributable to the spin Zeeman effect. The valley splitting slopes are different for the c1 and c2 bands due to the Zeeman effect, shown in Fig. \ref{fig:dos_magnetic}(c). Moreover, in the valence-band edge, both Landau levels and Zeeman splittings occur in the valence-band edge. 

\section{Conclusion}
 \label{conclusion}
 In this study, we evaluated the electronic properties of $\rm A\Tilde{A}B$ and $\rm A\Tilde{A}A$ both with and without relaxation. Our results indicate that the electronic properties of rigid $\rm A\Tilde{A}B$ are distinct from those of $\rm A\Tilde{A}A$. Before relaxation, the VB1 of rigid $\rm A\Tilde{A}B$ intersects with VB2 at the $K-$point, and it becomes isolated after relaxation, while $\rm A\Tilde{A}A$ has isolated and non-isolated VB1 before and after relaxation, respectively. The VB1 originates from the $\Gamma$-states of the monolayer. We analyzed the electron distributions of the two structures and found that an isolated VB1 is usually associated with the localization of electrons at high-symmetry points, which have the strongest interlayer interactions. Rigid $\rm A\Tilde{A}A$ have only one dominant high-symmetry stacking form in their interlayer interactions, leading to the localization of electrons and an isolated VB1. However, in rigid $\rm A\Tilde{A}B$, there are two high-symmetry stackings, AA and S-S, with strong interlayer interactions, and electrons are primarily distributed at these two regions, leading to a non-isolated VB1. We used this relationship between the interlayer interactions at high-symmetry points and electronic properties to open the gap between VB1 and VB2 in rigid $\rm A\Tilde{A}B$ by weakening the interlayer interaction of either the AA or S-S stacking. Due to the different symmetry, after introducing the SOC, we find a spin-valley-layer locking in the $\rm A\Tilde{A}B$, whereas a spin-valley locking in the $\rm A\Tilde{A}A$. We apply an electric field to open the gap in rigid $\rm A\Tilde{A}B$. The gap between VB1 and VB2 is opened through the synergistic effect of interlayer interactions and the electric field. Moreover, we found that the trend of the gap of $\rm A\Tilde{A}B$ changes differently from that of $\rm A\Tilde{A}A$ with the variation of the electric field $E$. In the $\rm A\Tilde{A}B$, due to the electric field, it decouples to an AA or AB tb-TMDs, opening a gap between VB1 and VB2 and changing the localization of the flat band states. Moreover, we examined how the energy levels of rigid $\rm A\Tilde{A}A$ respond to the magnetic field, both with and without SOC. We discovered that, in both scenarios, applying a magnetic field induces energy level splitting. This splitting is attributed to valley splitting, Zeeman shift, and Landau levels with quantized energies. Finally, compared with the tb-TMDs, we emphasize the peculiar properties observed in TTM. The TTM is more flexible and controllable, which is very sensitive to the starting stacking arrangements, lateral shift, and external fields. Two bilayer moir\'e patterns in the TTM have an interplay in both structural and electronic structures. The spin-valley-layer locking ensures that the spin in the TTM could be effectively manipulated by the electric field control of the layer polarization and magnetic field control of the valley polarization.
	
\section{Acknowledgments}
We thank Francisco Guinea, Pierre A. Pantale\'on, Adrian Ceferino and Xueheng Kuang for their useful discussions. This work was supported by the National Natural Science Foundation of China (Grant No. 11974263, 12174291) and the Knowledge Innovation Program of Wuhan Science
and Technology Bureau (Grant No. 2022013301015171). Z.Z. acknowledges support funding from the European Union's Horizon 2020 research and innovation programme under the Marie Skłodowska-Curie grant agreement No. 101034431, and from the ``Severo Ochoa" Programme for Centres of Excellence in R\&D (CEX2020-001039-S / AEI / 10.13039/501100011033).
We thank the Core Facility of Wuhan University for providing the computational resources.\\

\begin{figure*}[t]
    \centering    
    \includegraphics[width=\textwidth]{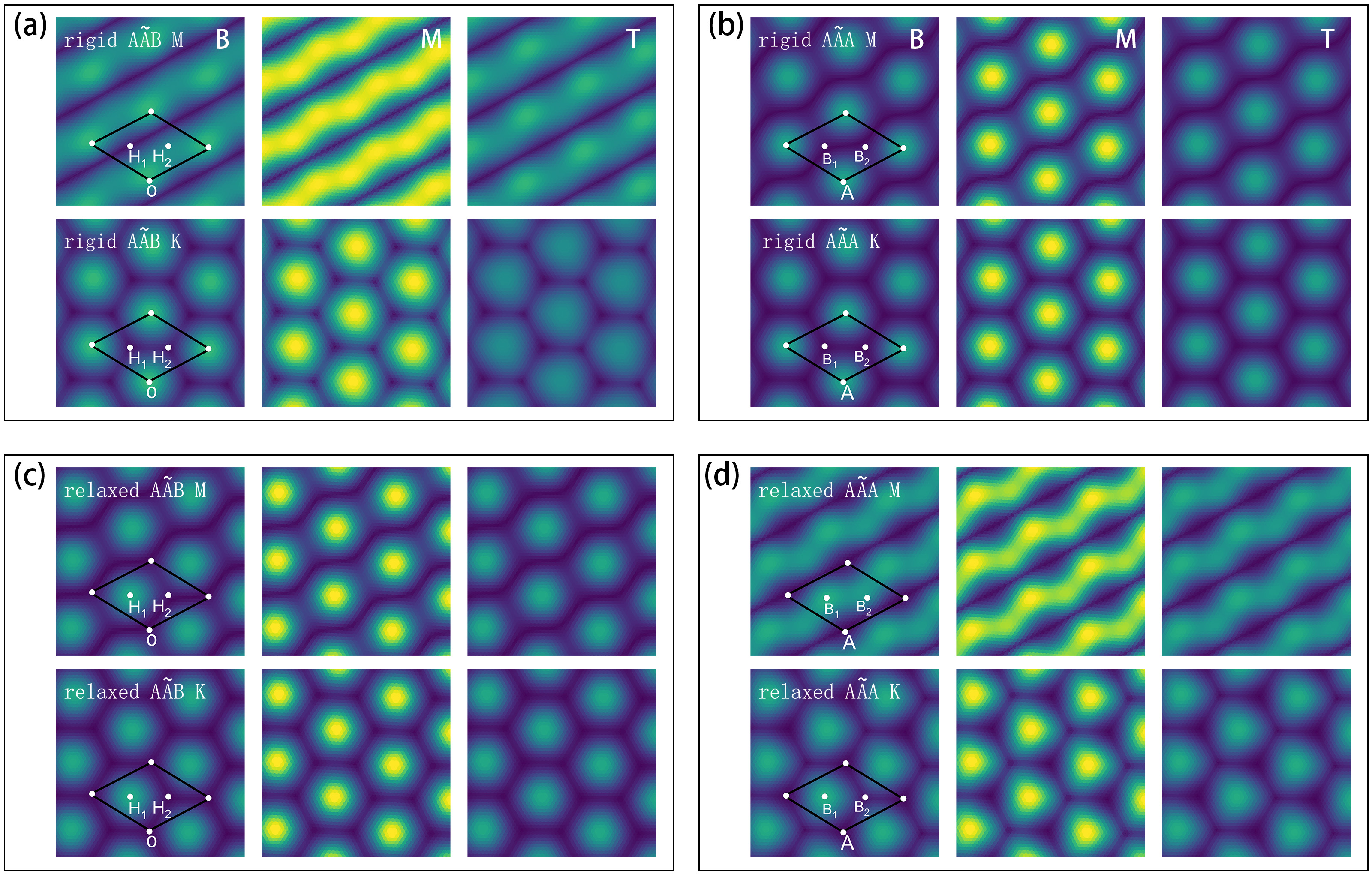}
    \caption{The eigenstates of the flat band at the $M-$ and $K-$point in $\rm A\Tilde{A}B-3.15^\circ$ and $\rm A\Tilde{A}A-3.15^\circ$. Parts (a) and (c) are eigenstates of rigid and relaxed $\rm A\Tilde{A}B$, and (b) and (d) are eigenstates of rigid and relaxed $\rm A\Tilde{A}A$, respectively. In each black rectangle, the panels from left to right are states in the bottom (B), middle (M), and top (T) layers, respectively. The unit cell and high-symmetry stackings are outlined in the bottom layer.}    \label{fig:eigenstates_appendix}
\end{figure*}
\appendix
\section{The eigenstates of the flat bands in $\rm A\Tilde{A}B$ and $\rm A\Tilde{A}A$ with and without relaxation}
\label{eigenstates_appendix}

\begin{figure}
    \centering    
    \includegraphics[width=0.5\textwidth]{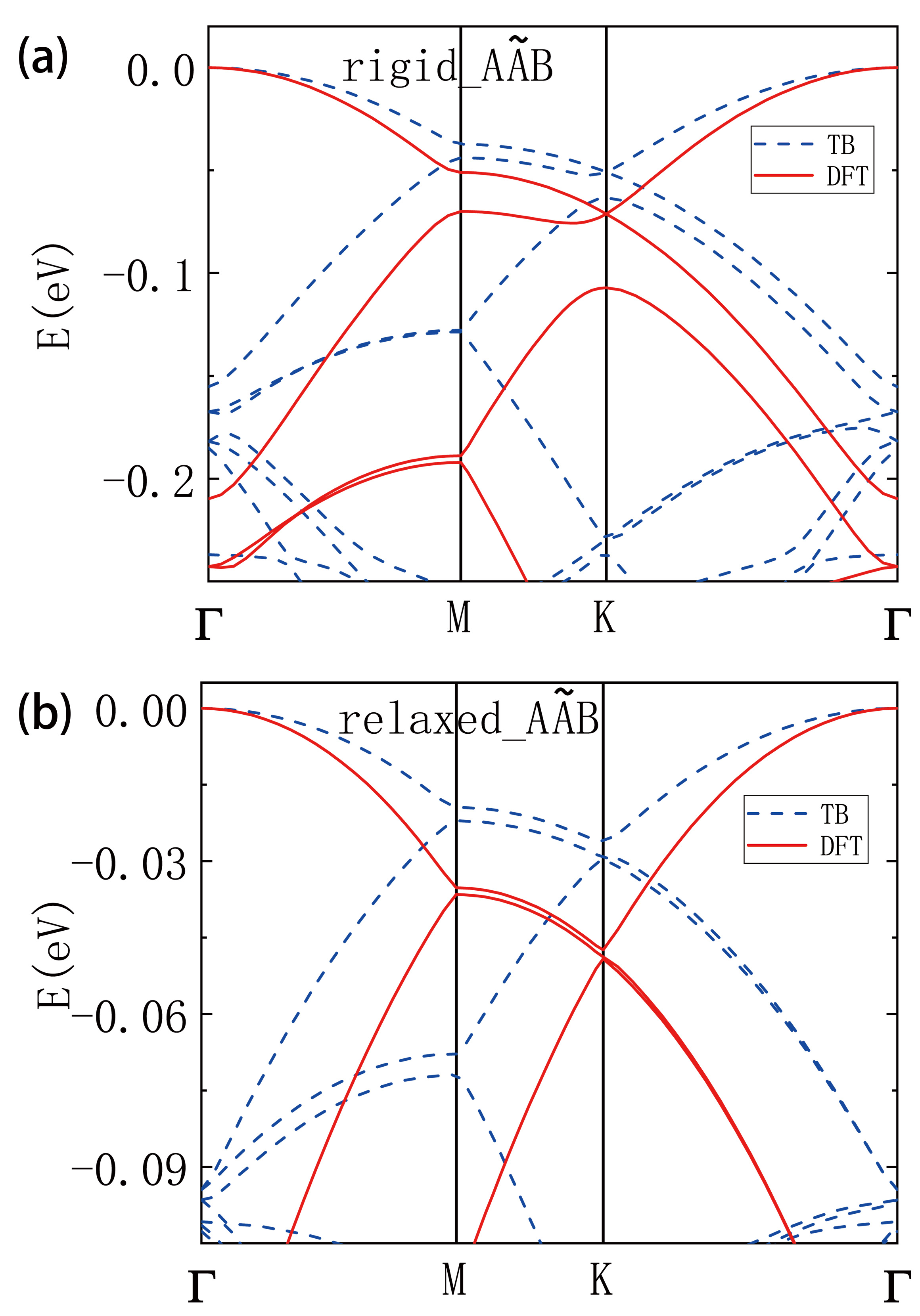}
    \caption{Comparison of rigid and relaxed DFT (solid red) vs TB (dashed blue) bandstructures at the twist angle, $\theta = 6.01^\circ$.}    \label{fig:tb_abacus}
\end{figure}

Figure \ref{fig:eigenstates_appendix} shows the eigenstates of $\rm A\Tilde{A}B$ and $\rm A\Tilde{A}A$ at $M-$ and $K-$point with and without relaxation. By comparing Fig. \ref{fig:eigenstates} and Fig. \ref{fig:eigenstates_appendix}, it is clear that the localization of the states of different $k$ points is the same in the isolated flat band case, but different in the non-isolated flat band case. 

\section{DFT vs TB band structures}
\label{dft_results}
In Fig. \ref{fig:tb_abacus}, we compare the DFT and TB band structures for rigid and relaxed $\rm A\Tilde{A}B$ at the twist angle, $\theta = 6.01^\circ$. DFT band structures have been computed with ABACUS\cite{Chen2010systematically, li2016large}. We use the Perdew-Burke-Ernzerhof exchange-correlation functional\cite{perdew1996generalized}, SG15 optimized Norm-Conserving Vanderbilt (ONCV) multi-projector pseudopotentials\cite{schlipf2015optimization} and standard atomic orbitals hierarchically optimized for the SG15-V1.0 pseudopotential \cite{lin2021strategy}. \textcolor{red}{In the relaxation process, the largest force among all of the atoms is 3.793e-3 eV/\AA, and the convergence criterion for self-consistent field (SCF) calculations during band-structure calculation is that the density difference between two SCF steps (DRHO) is less than 5e-8.} We found that the TB and DFT results have the same conclusion that VB1 touches VB2 at K-point for rigid $\rm A\Tilde{A}B$ and becomes an isolated band after relaxation, which indicates that our TB model can reach the same conclusion as DFT when studying the effect of relaxation on the band structures of TTM.

\section{The tight-binding model parameters}

\begin{table*}[htbp]
  \begin{center}
    \caption{The fitting parameters for interlayer hopping from Ref. \cite{fang2015abinitio}.}
    \label{fitting_parameters}
    \begin{tabular}{|c|c|c|c|c|c|c|c|c|c|c|}
        \hline
      a(nm) & c(nm) & d$_{S-S}$ & $\nu_\sigma$(eV) & R$_\sigma$(nm) & $\eta_\sigma$ & $\nu_\pi$(eV) & R$_\pi$(nm) & $\eta_\pi$ & $\lambda_{SO}^{Mo}$(eV/$\hbar^2$) & $\lambda_{SO}^{S}$(eV/$\hbar^2$) \\
      \hline
      0.316 & 0.6145 & 0.317 & 2.627 & 0.3128 & 3.859 & -0.708 & 0.2923 & 5.724 & 0.0836 & 0.0556 \\
      \hline
    \end{tabular}
  \end{center}
\end{table*}



\begin{table}[htbp]
    \begin{center}
    \caption{The interlayer distances (in nm) at high-symmetry points.}
    \label{interlayer_distances}
    \begin{tabular}{|c|c|c|c|c|c|c|}
    \hline
    & \multicolumn{3}{|c|}{relaxed $\rm A\Tilde{A}B$} & \multicolumn{3}{|c|}{relaxed $\rm A\Tilde{A}A$} \\
    \hline
     & O & H$_1$ & H$_2$ & A & B$_1$ & B$_2$\\
    \hline
    BM & 0.679 & 0.656 & 0.665 & 0.680 & 0.651 & 0.651\\
    \hline
    MT & 0.665 & 0.656 & 0.680 & 0.680 & 0.651 & 0.651\\
    \hline
    \end{tabular}
    \end{center}
\end{table}

\begin{table*}[htbp]
  \begin{center}
    \caption{The tight-binding hopping parameters for intralayer hopping.}
    \label{hopping_parameters}
    \begin{tabular}{|c|c|}
        \hline
        \thead{(0,0,0)} & 
        \thead{$\mathcal{E}_0=-0.1380$, $\mathcal{E}_1=\mathcal{E}_2=0.0874$,  $\mathcal{E}_3=\mathcal{E}_4=1.0688$, \\ $\mathcal{E}_5=\mathcal{E}_6=-1.5984$,  $\mathcal{E}_7=\mathcal{E}_10=-1.8352$, $\mathcal{E}_8=\mathcal{E}_9=-1.5984$, \\
        $t_{0,6}=t_{0,9}=-0.6648$, $t_{0,7}=-0.6248$, $t_{0,10}=0.6248$, $t_{1,5}=t_{1,8}=0.9980$, \\ $t_{2,6}=t_{2,9}=0.4608$, $t_{2,7}=-0.6742$, $t_{2,10}=0.6742$, $t_{3,5}=-0.5574$, $t_{3,8}=0.5574$, \\ $t_{4,6}=1.5262$, $t_{4,7}=t_{4,10}=-0.9751$,$t_{4,9}=-1.5262$, $t_{5,8}=t_{6,9}=-0.3082$, $t_{7,10}=1.0597$
        } \\
        \hline
        
        \thead{(1,0,0)} & 
        \thead{$t_{0,0}=-0.2979$, $t_{0,1}=0.4120$, $t_{0,2}=-0.1056$, $t_{0,5}=t_{0,8}=0.5758$, $t_{0,6}=t_{0,9}=0.3324$,\\ $t_{0,7}=-0.6248$, $t_{0,10}=0.6248$, $t_{1,1}=-0.3499$, $t_{1,2}=0.1119$, $t_{1,5}=t_{1,8}=0.5951$,\\ $t_{1,6}=t_{1,9}=t_{2,5}=t_{2,8}=-0.2326$, $t_{1,7}=0.5839$, $t_{1,10}=-0.5839$, $t_{2,2}=0.0665$, \\ $t_{2,6}=t_{2,9}=0.8637$, $t_{2,7}=0.3371$, $t_{2,10}=-0.3371$, $t_{3,3}=-0.0275$, $t_{3,4}=-0.3598$,\\ $t_{3,5}=1.0053$, $t_{3,6}=t_{4,5}=0.9022$, $t_{3,7}=t_{3,10}=0.8444$, $t_{3,8}=-1.0053$, \\ $t_{3,9}=t_{4,8}=-0.9022$,  $t_{4,4}=-0.1471$, $t_{4,6}=-0.0365$, $t_{4,7}=t_{4,10}=0.4876$, \\ $t_{4,9}=0.0365$,  $t_{5,5}=t_{8,8}=0.1542$, $t_{5,6}=t_{8,9}=0.3695$, $t_{5,7}=-0.0244$, \\ $t_{5,8}=0.0783$, $t_{5,9}=-0.0156$, $t_{5,10}=-0.0841$, $t_{6,6}=t_{9,9}=0.6438$, $t_{6,7}=0.0180$, \\ $t_{6,8}=-0.0476$, $t_{6,9}=0.0418$, $t_{6,10}=-0.0703$, $t_{7,7}=t_{10,10}=-0.1828$, \\ $t_{7,8}=-0.0188$, $t_{7,9}=-0.1080$, $t_{7,10}=0.0088$, $t_{8,10}=0.0244$, $t_{9,10}=-0.0180$
        } \\
        \hline

        \thead{(0,1,0)} & 
        \thead{$t_{0,0}=-0.2979$, $t_{0,1}=-0.4120$, $t_{0,2}=-0.1056$, $t_{0,5}=t_{0,8}=-0.5758$, $t_{0,6}=t_{0,9}=0.3324$,\\ $t_{0,7}=-0.6248$, $t_{0,10}=0.6248$, $t_{1,1}=-0.3499$, $t_{1,2}=-0.1119$, $t_{1,5}=t_{1,8}=0.5951$,\\ $t_{1,6}=t_{1,9}=t_{2,5}=t_{2,8}=0.2326$, $t_{1,7}=-0.5839$, $t_{1,10}=0.5839$, $t_{2,2}=0.0665$, \\ $t_{2,6}=t_{2,9}=0.8637$, $t_{2,7}=0.3371$, $t_{2,10}=-0.3371$, $t_{3,3}=-0.0275$, $t_{3,4}=0.3598$,\\ $t_{3,5}=1.0053$, $t_{3,6}=t_{4,5}=-0.9022$, $t_{3,7}=t_{3,10}=-0.8445$, $t_{3,8}=-1.0053$,\\ $t_{3,9}=t_{4,8}=0.9022$, $t_{4,4}=-0.1471$, $t_{4,6}=-0.0365$, $t_{4,7}=t_{4,10}=0.4876$, \\ $t_{4,9}=0.0365$, $t_{5,5}=t_{8,8}=0.1542$, $t_{5,6}=t_{8,9}=-0.3695$, $t_{5,7}=0.0244$, \\ $t_{5,8}=0.0783$,  $t_{5,9}=0.0156$,  $t_{5,10}=0.0840$, $t_{6,6}=t_{9,9}=0.6438$, $t_{6,7}=0.0180$, \\ $t_{6,8}=0.0476$, $t_{6,9}=0.0418$, $t_{6,10}=-0.070$, $t_{7,7}=t_{10,10}=-0.1828$,  $t_{7,8}=0.0188$, \\ $t_{7,9}=-0.1080$, $t_{7,10}=0.0088$, $t_{8,10}=-0.0244$, $t_{9,10}=-0.0180$
        } \\
        \hline

        \thead{(1,1,0)} & 
        \thead{$t_{0,6}=t_{0,9}=-0.1059$, $t_{0,7}=-0.0485$, $t_{0,10}=0.0485$, \\ $t_{2,6}=t_{2,9}=-0.1733$, $t_{2,7}=-0.1559$, $t_{2,10}=0.1559$
        } \\
        \hline

        \thead{(1,-1,0)} & 
        \thead{$t_{0,0}=-0.2979$, $t_{0,1}=0.1145$, $t_{0,2}=0.4096$, $t_{0,5}=t_{0,8}=-0.0917$, $t_{0,6}=t_{0,9}=0.0529$,\\ $t_{0,7}=-0.0485$, $t_{0,10}=0.0485$, $t_{1,1}=0.2747$, $t_{1,2}=0.2487$, $t_{1,5}=t_{1,8}=-0.1300$, \\ $t_{1,6}=t_{1,9}=t_{2,5}=t_{2,8}=0.0750$, $t_{1,7}=-0.1350$, $t_{1,10}=0.1350$, $t_{2,2}=-0.5581$,\\ $t_{2,6}=t_{2,9}=-0.0433$, $t_{2,7}=0.0780$, $t_{2,10}=-0.0780$, $t_{3,3}=-0.2069$, $t_{3,4}=0.2562$,\\ $t_{4,4}=0.0323$, $t_{5,5}=t_{8,8}=0.8886$, $t_{5,6}=t_{8,9}=0.0545$, $t_{5,7}=0.0034$, $t_{5,8}=0.0236$,\\ $t_{5,9}=-0.0160$, $t_{5,10}=t_{7,8}=-0.1029$, $t_{6,6}=t_{9,9}=-0.0906$, $t_{6,7}=-0.0302$,\\ $t_{6,8}=0.0160$, $t_{6,9}=0.0966$, $t_{6,10}=-0.0377$, $t_{7,7}=t_{10,10}=-0.1828$, \\$t_{7,9}=0.0377$, $t_{7,10}=0.0088$, $t_{8,10}=-0.0034$, $t_{9,10}=0.0302$
        } \\
        \hline
    \end{tabular}
  \end{center}
\end{table*}

In Tab. \ref{fitting_parameters}, a is the lattice constant, c is the interlayer distance, d$_{S-S}$ is the nearest distance between S atoms in out-of-plane direction,$\lambda_{SO}^{Mo/S}$ is the atomic spin-orbit coupling strength, and the remaining parameters are the interlayer hopping calculation formula parameters used in Eq. \ref{equation:interlayer_hopping}. For monolayer MoS$_2$, the tight-binding hopping parameters are present in Tab. \ref{hopping_parameters}, Combining the TTM structural parameters and interlayer hopping parameters in Tab. \ref{fitting_parameters}, we can use Eq. \ref{equation:interlayer_hopping} to calculate the interlayer hopping terms of TTM. Then, by combining the intralayer hopping parameters in Tab. \ref{hopping_parameters}, we can use TBPLaS to calculate the electronic properties of TTM\cite{li2023tbplas}. \textcolor{red}{In Tab. \ref{interlayer_distances}, we list the interlayer distances at high-symmetry points of relaxed $\rm A\Tilde{A}B$ and relaxed $\rm A\Tilde{A}A$. The interlayer distances are all between 0.65 nm and 0.68 nm, with the largest interlayer distances found in AA stacking and S-S stacking, reaching up to 0.68 nm. Ab initio calculations (vdW-DF) yield interlayer distances of 0.677 nm for AA stacking and 0.623 nm for AB stacking\cite{he2014stacking}. Compared to the rigid structures of $\rm A\Tilde{A}B$ and $\rm A\Tilde{A}A$, the relaxed structures exhibit interlayer distances that are closer to the realistic case. What's more, the study revealed that when changes in intralayer hopping, caused by atomic displacements within layers, are disregarded, the differences in electronic properties between rigid and relaxed twisted bilayer MoS$_2$ are similar to our conclusions\cite{vitale2021flat}. Therefore, we can infer that the primary source of difference in electronic properties between rigid and relaxed structures in TTM (Figs. \ref{fig:electric}d and h) mainly arises from variations in the interlayer spacing.}
\bibliography{bibliography.bib}
\end{document}